# On the reinforcement homogenization in CNT/Metal Matrix Composites during Severe Plastic Deformation

Katherine Aristizabal[a], Andreas Katzensteiner[b], Andrea Bachmaier[b], Frank Mücklich[a], Sebastian Suárez[a]

[a]Chair of Functional Materials, Department of Materials Science, Saarland University, Campus 66123, Saarbrücken, Germany

[b]Erich Schmid Institute of Materials Science, Austrian Academy of Sciences, Jahnstrasse 12, A-8700 Leoben, Austria

Corresponding author. katherine.aristizabal@uni-saarland.de

**Abstract**

Carbon nanotube (CNT)-reinforced nickel matrix composites with different concentrations were processed by high pressure torsion (HPT). We thoroughly characterized the CNT agglomerates' spatial arrangement at different stages of deformation in order to extract information valuable for the optimization of the processing parameters and to elucidate the mechanisms involved during the processing of particle reinforced metal matrix composites by HPT. From the electron micrographs taken on the radial direction with increasing equivalent strains, we observed that CNT agglomerates debond by relative sliding between CNT during HPT, becoming spherical at higher stages of deformation. Furthermore, we introduced a model for the prediction of the minimum strain required for a homogeneous distribution of a second phase during HPT, which can be correlated to the material's three-dimensional structure and agrees well with the experimental data.

# 1. Introduction

The distribution of reinforcing phases in composite materials is of great importance and has a big influence on their mechanical performance. Carbon nanotubes have been widely used as reinforcing phase not only in polymer [1,2] and ceramic matrix composites [3,4] but also in metal matrix composites [5–7]. Some authors have explored different strategies for improving the distribution of CNT in MMC such as blending by mixing, nano-scale dispersion, ball milling, cold spraying and molecular-level mixing [6,8,9] However, blending by mixing has been found to deteriorate the mechanical properties of the composites due to poor distribution of the CNT. Although nano-scale dispersion was found to improve significantly the distribution of the CNT in Aluminum matrix, because it consist in utilizing natural rubber in a mixture with the CNT and the metallic powder alternatively stacked in a preform, it needs to be subjected to high temperatures (800°C) in order to burn the rubber and to melt the metal, which would imply the use of even higher temperatures in the case of nickel, which has a melting point of 1455°C, adding further difficulties to the manufacturing process. Ball milling and molecular-level mixing methods also improve significantly the dispersion of the CNT, both producing large agglomerates' sizes of about some microns to some millimeters. Ball milling also results in severe damage to the CNT. A thorough discussion of the advantages and drawbacks of each technique is beyond the scope of this manuscript and can be found in [6] and the references therein. Therefore, given their tendency to form agglomerates due to Van der Waals forces, a homogeneous distribution of CNT is still a challenging task.

High-pressure torsion, a severe plastic deformation process, has shown to be a powerful tool for improving the distribution of particles in MMC. I. Sabirov et al, showed how ceramic particles are reduced in size and dispersed inside the metallic matrix by debonding [10]. During HPT, the sample is placed between two anvils using high pressures (> 2 GPa) and rotated in quasi-constrained conditions a certain number of turns (T). By the action of plastic flow and the generation, mobility and re-arrangement of dislocations, a refinement of the microstructure takes place and the strength of the material increases with increasing strain until the saturation in the microstructural refinement is reached [11]. Nevertheless, these microstructures possess high stored energy in their large grain boundary area. Recently, based on their ability to pin the microstructure [5], CNT have been used as stabilizing phase against grain growth [12,13] in MMC processed by severe plastic deformation. However, in order to efficiently fulfill their stabilization task, CNT should be homogeneously distributed [14], which is also expected to increase the mechanical performance of these MMC. For that purpose, the processing route used in this work started by colloidal mixing of the CNT with the metallic Nickel powder, followed by the cold pressing and hot sintering of the blends and succeeding processing by means of high pressure torsion.

For qualitative assessment of the distribution, directly observation of electron micrographs is sufficient. Previously, a preliminary assessment of the reinforcement homogeneity in CNT/Ni MMC was carried out [15]. In the present study the aim is to extract quantitative information from the electron micrographs on the CNT agglomerate size and spatial distribution in MMC subjected to severe plastic deformation, and correlate it to the mechanisms involved in the deformation of the reinforcement during processing, seeking the optimization of the process parameters and the improvement of the physical properties of the composites. Some quantitative methods on the evaluation of the dispersion and distribution of CNT in MMC have been proposed in the literature [16–19]. A thorough discussion of the different quantification methods can be found elsewhere [20]. Furthermore, event-to-event (nearest neighbor distance NND), based methods, commonly used in spatial statistics, have been used in unidirectional composites [21]. Moreover, the method of Region Homogeneity $H_{RO}$ has been proposed as an easy way of assessing the distribution homogeneity of second phases in metallic materials [22] and also of CNT agglomerates in MMC [23]. The advantages and limitations of the latter in the evaluation of MMC processed by SPD are discussed here.

In this work, an extensive clustering and distribution homogeneity analysis was performed with the aim of understanding the behavior of the CNT during the deformation process as a function of the accumulated deformation. For the evaluation of the studied composites' homogeneity, a NND clustering analysis-based methodology is used, which can be completely carried out using open source software [24,25]. From the electron micrographs, a CNT agglomerate-debonding mechanism by relative sliding between CNT during HPT is observed, which contributes to the understanding of the reinforcement arrangement after the processing by severe plastic deformation of CNT MMC. Finally, a model for the prediction of the minimum equivalent strain that should be applied during HPT for a homogenous distribution of particles in MMC is proposed, which provides a basis for the optimization of the processing parameters.

## 2. Experimental

2.1. Manufacturing and HPT processing of CNT/Ni composites

The composites were obtained via powder metallurgy and further processed by HPT. The starting materials were MWCNT (CCVD grown, Graphene Supermarket, USA density 1.84 g/cm$^3$, diameter: 50-85 nm, length: 10-15 μm, carbon purity: >94%) and dendritic Ni powder (Alfa Aesar, mesh -325). A colloidal mixing process was used to blend the precursor powders by which CNT are dispersed in ethylene glycol EG (CNT/EG concentration ratio at 0.2 mg/ml) and mixed with Ni powder. A thorough description of this process is reported elsewhere [26]. The CNT fractions used were 0.5, 1, 2 and 3 wt. % (2.4, 4.7, 9 and 13 vol. %, respectively). The powders were dried and cold pressed under 990 MPa and subsequently sintered in a hot uniaxial press HUP under vacuum (2 x 10$^{-6}$ mbar) at 750 °C for 2.5 h with a 264 MPa axial pressure. Sintered samples were further processed by means of HPT at room temperature using 1, 4, 10 and 20 T under 4 GPa of pressure. Samples with 1 wt. % CNT were processed 30 T at room temperature RT and at 200ºC, were also analyzed. Fig. 1 shows schematically the HPT set up.

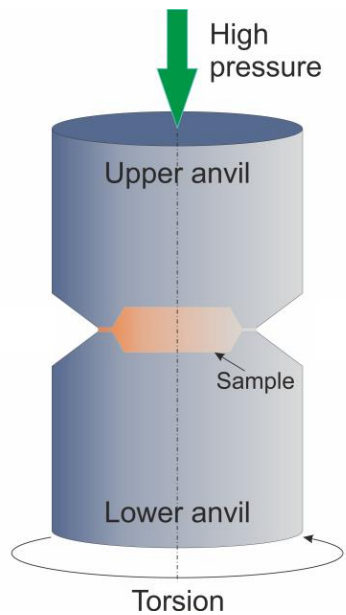

**Figure 1 Schematic HPT set up used for sample processing.**

2.2. Characterization

The HPT samples were cut in halves, embedded in conductive resin and fine polished using polishing discs with the aid of 6, 3 and 1 μm diamond suspensions and finally with OPS colloidal silica. The samples were then characterized by scanning electron microscopy (SEM) using a Helios NanoLab™ 600 dual beam field emission microscope (FEI Company). The analysis of the microstructure was carried out along the middle plane, in order to avoid the effect of the microstructural gradients along the height of the specimens [27]. The images were taken every 1 mm with a resolution of 12.5 nm per pixel along the radial direction, corresponding to increasing equivalent strain values, according to $\varepsilon_{eq} = \frac{2\pi T r}{t\sqrt{3}}$, where T is the number of turns, t is the sample thickness and $r$ the distance from the center of the sample [28]. In the case of HPT samples, the micrographs were acquired at 10 kX magnification, resulting in a field of view of (12.8 x 12.8) μm². The choice was made based on the size of the agglomerates: if the images were to be taken at lower magnifications, information about the smaller agglomerates in the case of the highly deformed samples would be lost due to resolution issues. On the other hand, at higher magnifications the larger agglomerates would be also neglected, since they would hardly fit into the region of interest. Furthermore, the size was kept constant for comparison purposes.

The images were digitally binarised and analyzed using the image processing package FIJI [24]. Micrographs with corresponding binary images are shown exemplarily in Fig. 2. The CNT agglomerates were considered as compact particles and a thorough particle analysis was performed on each image. The particle analysis protocol was as follows: the images were calibrated with the known scale; the threshold was adjusted thoroughly, in order to separate the dark (CNT agglomerates) from the light regions (Ni matrix), without removing pixels from the boundaries (manual segmentation was performed when necessary) and finally, the images were made binary. Different size descriptors (such as the area, the maximum and minimum Feret diameter and the perimeter) and shape factors (such as roundness, circularity, aspect ratio etc.) can be obtained during particle analysis. In this case, the particle area was obtained and the diameter of an equivalent circumference was computed as the agglomerate equivalent diameter $D_{CNT}$. Additionally, the position of the agglomerate centers of mass was extracted and used to calculate the NND (nearest neighbor distance). The evolution of the area weighted agglomerate diameter and the mean NND with increasing equivalent strain were studied (Fig. 3).

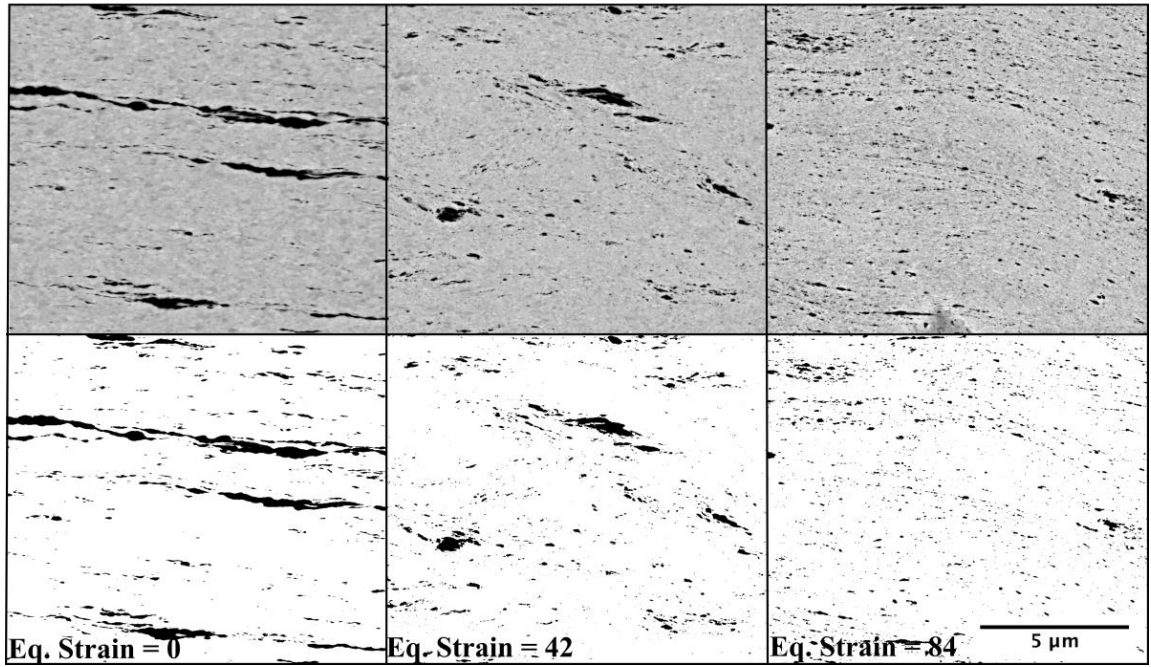

**Figure 2 Electron micrographs (upper row) and their respective binarized images (lower row) from a 1 wt.% sample after 10 turns. The equivalent strain increases from left to right. The dark regions correspond to CNT agglomerates.**

2.3 Quantitative assessment of the distribution homogeneity and clustering behavior

Furthermore, a set of samples was analyzed in terms of homogeneity and distribution using two different methodologies. The quantitative assessment of the homogeneity was carried out following the method proposed by Rossi et al [23]. The agglomerate area fraction (phase amount) and the number of objects were obtained using particle analysis in 20 different regions of interest using ROI manager in FIJI. The region homogeneity $H_R$ is a constructed homogeneity obtained by multiplying the object number homogeneity $H_{NO}$ and the phase amount homogeneity $H_{PA}$. The partial homogeneities H were calculated as the counterpart of the Gini G coefficients G: $H = 1 - G$, which is a measure of distribution inequality [22]. The Gini index of each parameter (object number and phase amount as obtained from particle analysis in FIJI) was retrieved using the *ineq* package in R studio, an integrated development environment for R, which is a programming language for statistical computer and graphics [25].

Information about the agglomerates' centers of mass was used for further point pattern analysis by means of the nearest neighbor distance distribution function (G-function) using the *spatstat* package in R Studio [29], in order to obtain more detailed information about the spatial behavior between the individuals (CNT agglomerates' centers of mass, treated as "points") of the investigated population (studied areas, treated as "point patterns"). This method consists on the comparison of the empirical cumulative distribution function $G_{obs}(r)$, with the nearest neighbor distribution function $G_{theo}(r)$ for complete spatial randomness (CSR). The nearest neighbor distance distribution function for a CSR is described by $G_{theo}(r)=1-\exp(-N_A\pi r^2)$, where $N_A$ is the number of particles per unit area, and r the evaluated distance. According to this, the condition $G_{obs}(r) < G_{theo}(r)$ is inherent of regular patterns and the opposite case, $G_{obs}(r) > G_{theo}(r)$, is interpreted as "clustering" (for example, Fig. a-c) because the NND are smaller than expected for the CSR case [30–32]. As there are numerous configurations for CSR, pointwise critical envelops with fixed respective number of points (corresponding to each case), obtained from 100 Monte Carlo simulations of CSR with a significance level of 2/101 = 0.0198 and spatial Kaplan-Meier edge correction, which is performed in order to correct effects arising from the non-visibility of points lying outside the evaluated field of view during the evaluation of the G-Function [33], were plotted as the theoretical expected behavior.

Accordingly, the G-function is a useful tool in spatial statistics that summarizes the "clustering" behavior of points in a point pattern. It displays the empiric cumulative distribution of NND in contrast to that of the CSR with the same number of points within a window of the same size. If a point pattern does not adjust to a Poisson pattern, then it can be either *clustered* (i.e. when the particles are interacting and tend to come closer together) or *regular* (i.e. when the points tend to avoid each other). It might be inferred that a homogeneous sample, e.g. with a high $H_{RO}$ value, also does not display a clustering behavior and vice versa. The G-Function can help to confirm or deny this statement. In this case, border corrections are carried out and the size of the window is arbitrary. For this reason, the G-function was chosen as a complementary method.

**3. Results and discussion**

By plotting the area weighted mean agglomerate diameter $D_{CNT}$ and the mean NND vs. Eq. Strain (Fig. 3) the evolution of both parameters with increasing strain can be tracked. Results show that both, $D_{CNT}$ and NND decrease significantly in size during the first stages of deformation approximately up to a strain of 20, but do not significantly change afterwards, where both $D_{CNT}$ and NND are between 100 and 400 nm. The deviation of the data also decreases with increasing strain. Fig. 3b also suggests that the agglomerates come closer for higher CNT concentration. Furthermore, a previous study showed that the agglomerate area fraction does not change significantly with increasing strain values [34]. Fig. 3 also displays information for a sample with 1 wt. % CNT processed with 30 T at 200 ºC. In the latter case, $D_{CNT}$ and NND are also within the same range discussed before.

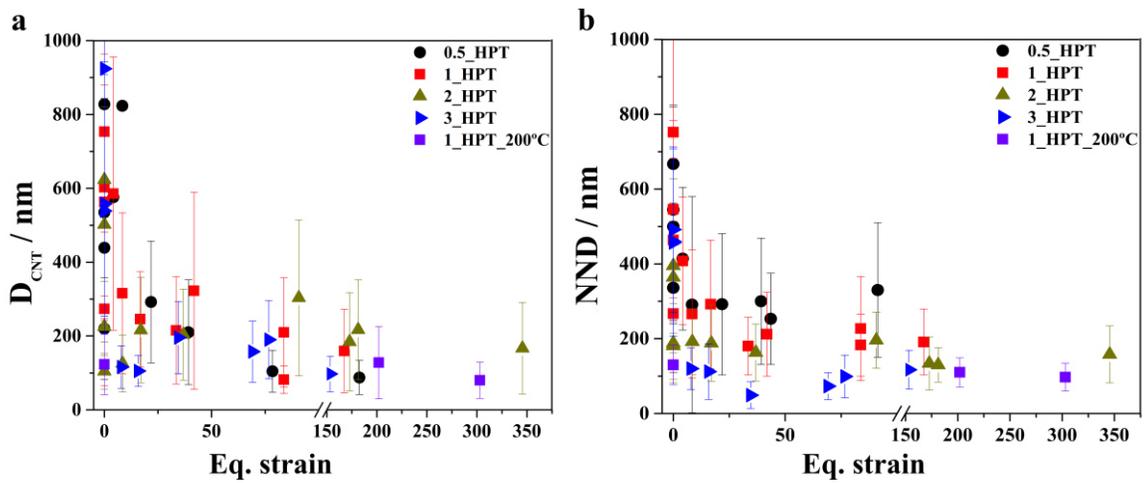

**Figure 3 a) Evolution of agglomerate diameter $D_{CNT}$, and b) evolution of the mean inter-particle distance (defined as NND) with increasing strain, of samples with different compositions.**

Nevertheless, from these results, an explicit relation concerning the evolution of the particle homogeneity during HPT is not feasible. In order to address this matter, analysis of the distribution of the CNT was carried out using the region homogeneity parameter $H_{RO}$ and the nearest neighbor distribution function G.

The parameter region homogeneity $H_{RO}$, as proposed by Rossi et al. [23] serves as a general evaluation of the distribution of particles lying within a studied area divided in an arbitrary number of regions, by quantifying the similarity of the different regions in terms of number of objects and amount of phase. This method is a practical and straightforward way of assessing the homogeneity distribution, when studying well-dispersed particle composites. Nevertheless, it does not take into account the separation or the interaction of the agglomerates (it is not sensitive to the clustering behavior of particles within the analyzed regions). Furthermore, the analysis is limited to samples (micrographs) of the same dimensions, since this method is area size sensitive.

On the other hand, when studying non-deformed MMC samples, lower magnifications can be used in order to display a higher number of particles without neglecting any significant information since the initial agglomerates are much larger ($D_{CNT} > 2\mu m$) than in the case of highly deformed samples ($D_{CNT}$ ~100 nm, Fig. 3). Furthermore, it is possible to keep the size of the analyzed area constant and $H_{RO}$ can be used to compare the homogeneity of different kind of MMC samples. Nevertheless, the inability of using lower magnifications and of keeping the size of the micrographs constant without losing significant information restricts the usefulness of $H_{RO}$ in highly deformed samples.

Fig. 4d displays the evolution of $H_{RO}$ for the samples with 1 wt. % CNT processed at RT. Even though $H_{RO}$ does not significantly improve during HPT up to 30 T, in most cases it is true that $H_{RO}$ increases with increasing strain except after only 1 T. Also, the diameter of the agglomerates and the inter-particle distance decrease significantly compared to the HUP samples (Fig. 3), and this is also an indicative of the improvement of the agglomerate distribution. Fig. 4a to 4c correspond to the G-function of samples with 1 wt. % CNT processed 10 T, which displayed the higher improvement in $H_{RO}$ (Fig. 4d). In all the cases a clustered behavior is present, as the empiric distributions are above the envelope for 100 simulations of the theoretical CSR expectation. In the case of samples with 1 wt. % CNT processed 30 T, $H_{RO}$ increases slightly with increasing strain above $H_{RO} = 60$ % (Fig. 4d). Fig. 4e to 4h show the G-function for the latter case. It can be seen from Fig. 4e that even for zero equivalent strain the empiric cumulative distribution of NND is closer to the CSR envelope. Nevertheless, for distances > 50 nm, a slight clustering behavior is observable. It can be inferred that for eq. strains $84 < \varepsilon < 100$ (Fig. 4c and 4f) the agglomerate NND cumulative distribution starts to behave homogeneously (Fig. 4f to 4h $G_{obs}$ stays within the envelope for completely random Poisson distributions).

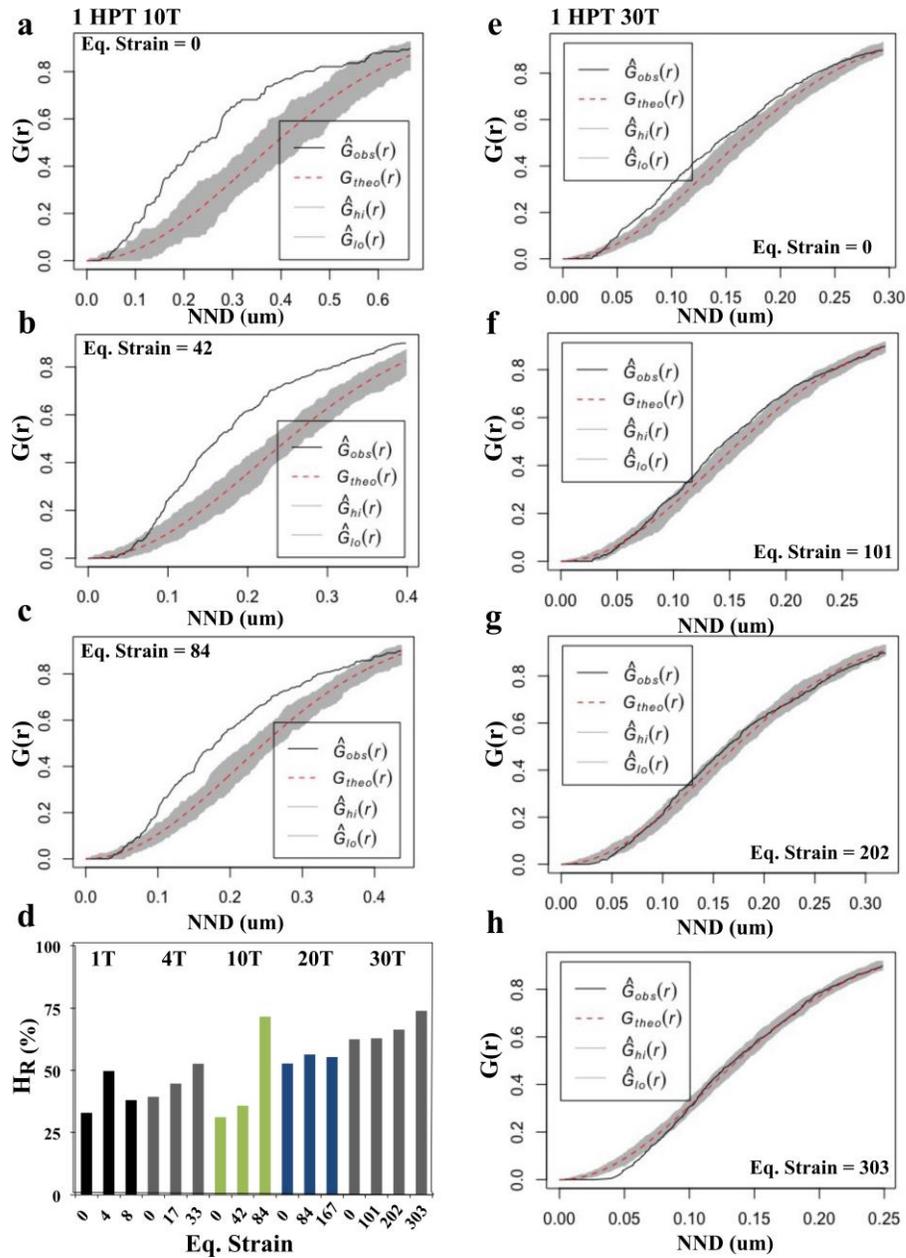

**Figure 4 (a-c) G(r) at different equivalent strains for 1wt.% CNT deformed at 10 T. (d) $H_{RO}$ for the selected set of samples. (e-f) G(r) at different equivalent strains for 1wt.%CNT deformed at 30 T. Where $\hat{G}_{Obs}$ (r) is the observed value of G(r); $G_{theo}$ (r) is the theoretical value of G(r) for complete spatial randomness; $G_{lo}$ (r) and $G_{hi}$ (r) represent the lower and upper bounds of G(r) from simulations. The deformation was performed at room temperature.**

A similar behavior can be seen in Fig. 5 for the samples with 1 wt. % CNT processed 30 T at 200ºC. In this case is also true that when ε = 0, G(r) is closer to the theoretical expectation CRS and for higher ε, G(r) is within the CRS envelopes, and for distances < 100 nm the agglomerates are more separated than the CRS case.

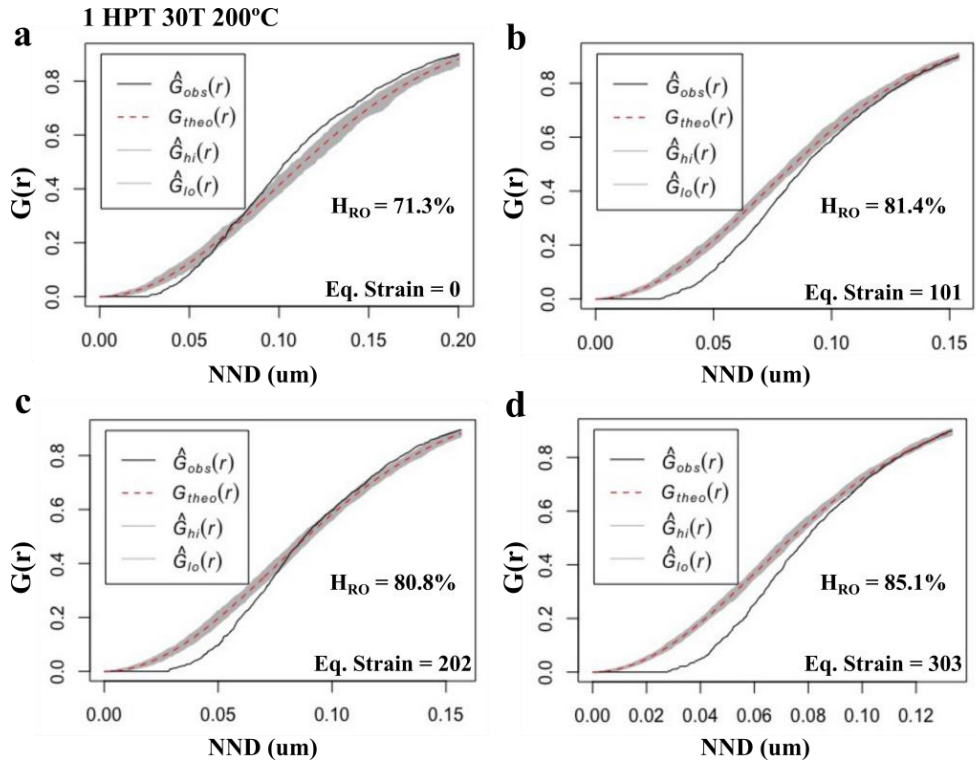

**Figure 5 G(r) for a 1wt.%CNT sample processed with 30 T at 200ºC, as a function of the equivalent strain. The measured $H_{RO}$ is also displayed.**

Sabirov [10] proposed the extended Tan and Zhang model (originally conceived for cold rolled MMCs) as an assessment of the equivalent strain values required in HPT to achieve a homogeneous particle distribution in ceramic particle reinforced MMC subjected to extrusion according to $d_p \geq \dfrac{d_m}{\left[\left(\frac{\pi}{6f}\right)^{\frac{1}{3}}\right]\sqrt{R}\,\gamma}$ where $d_p$ is the particle size, $d_m$ is the matrix powder size required to attain a homogeneous particle distribution; f the particle volume fraction; R is the extrusion ratio; γ is the shear strain with $\gamma = \sqrt{3}\varepsilon_{eq}$. This is based on the assumption that the particles are spherical. They found that SiC and Al$_2$O$_3$ particles behave differently during HPT and that the homogenization occurs according to Tan and Zhang model [35]. Nevertheless, the experimental data did not adjust well to their proposed model, which was attributed to the fact that the ceramic particles are de-clustered through a debonding mechanism without deformation of the particle clusters.

Although, in the samples studied here, the agglomerates cannot be considered spherical throughout the entire process because during the first stages of deformation elongated CNT agglomerates in the shear direction are observed (Fig. 2, ε = 0). Nevertheless, at higher applied strains, they start to debond forming more spherical agglomerates (Fig. 2, eq. ε = 84). Contrary to ceramic particles, which are hard and brittle, CNT are elastic and CNT agglomerates are bonded by Van der Waals forces that can be overcome during deformation. According to this, CNT agglomerates debond by relative sliding between CNT. Furthermore, sliding between CNT walls may also occur [36].

In the literature, a unique value of inter-particle distance, which only deals with the case of equidistant particles, has been considered for the assessment of a uniform distribution of particles during plastic deformation [10,35]. Nevertheless, as already discussed, the particles can be homogeneously distributed according to different configurations describing a Poisson distribution. Therefore, a model is proposed, starting from the nearest neighbor distribution function in 3D, and its correlation with the CNT agglomerates' spatial behavior. In this case, HPT is considered and the agglomerates are assumed to be spherical.

The nearest neighbor distance NND distribution function in 3D for completely random Poisson distribution is described by equation 1 [37]:

$$D(r) = 1 - \exp\left(-N_V \frac{4\pi}{3} r^3\right); r \geq 0 \qquad (1)$$

where $N_V$ is the mean number of objects per unit volume. $N_V$ for isolated spherical objects of random diameter D is [37] (pg. 78):

$$N_V = \frac{6 V_V}{\pi D^3} \qquad (2)$$

Replacing 2 in equation 1 yields:

$$D(r) = 1 - \exp\left(-\frac{8 V_V}{D^3} r^3\right) \qquad (3)$$

For point fields it is true that the volume fraction equals the spherical contact distribution $V_V = H^s(r)$ and from the completely random property of the Poisson field it follows that the NND distribution $D(r)$ and the spherical contact distribution function $H^s(r)$ are identical [37] (pg. 315) $H^s(r) \equiv D(r)$, thus $D(r) = V_V$. According to this and solving r in equation 2:

$$r = \frac{D}{2}\left[-\frac{\ln(1-V_V)}{V_V}\right]^{1/3} \qquad (4)$$

Also, for uniform distribution of CNT, which are expected to be at the grain boundaries due to the nature of the processing route (powder metallurgy), r should be equal or greater than the grain size of the matrix after HPT $GS_{fHPT}$:

$$r \geq GS_{fHPT} \qquad (5)$$

Furthermore, for simple shear a volume element, which may be a grain or a second phase in a composite, will be deformed in an ellipsoid with apex ratio $r'$. The reduction ratio of an ellipsoid is given by $\sqrt{r'}$. For large shear strains ($\gamma > 2$), $r' = \gamma^2$, with $\gamma = \sqrt{3}\varepsilon_{eq}$ [28]. In the case of the studied CNT reinforced MMC, it is assumed that the agglomerates are deformed in the same way as the matrix. Accordingly, after HPT the reduction ratio is given by:

$$\frac{GS_{iHPT}}{GS_{fHPT}} = \sqrt{3}\varepsilon_{eq} \tag{6}$$

$GS_{iHPT}$ corresponds to the grain size of the matrix before HPT, i.e. the grain size of the composite after the sintering process by HUP $GS_{iHPT} = GS_{fHUP}$. Thus, replacing in equation 6 and solving $GS_{fHUP}$:

$$GS_{fHPT} = \frac{GS_{fHUP}}{\sqrt{3}\varepsilon_{eq}} \tag{7}$$

and equation 5 can be expressed as:

$$\frac{D}{2}\left[-\frac{\ln(1-V_V)}{V_V}\right]^{1/3} \geq \frac{GS_{fHUP}}{\sqrt{3}\varepsilon_{eq}} \tag{8}$$

Previously, it was shown for consolidated CNT reinforced Nickel matrix composites that, in the presence of CNT a grain growth stagnation takes place during sintering and eventual annealing and the final grain size is related to the CNT volume fraction $V_V$ according to the Zener based relationship $GS_{fHUP} = \frac{0.99 \pm 0.07}{V_V^{0.4}}$ [38].

Fig. 6 shows the experimental data of the mean grain size measured by EBSD of the studied composites before HPT with the corresponding Zener based model.

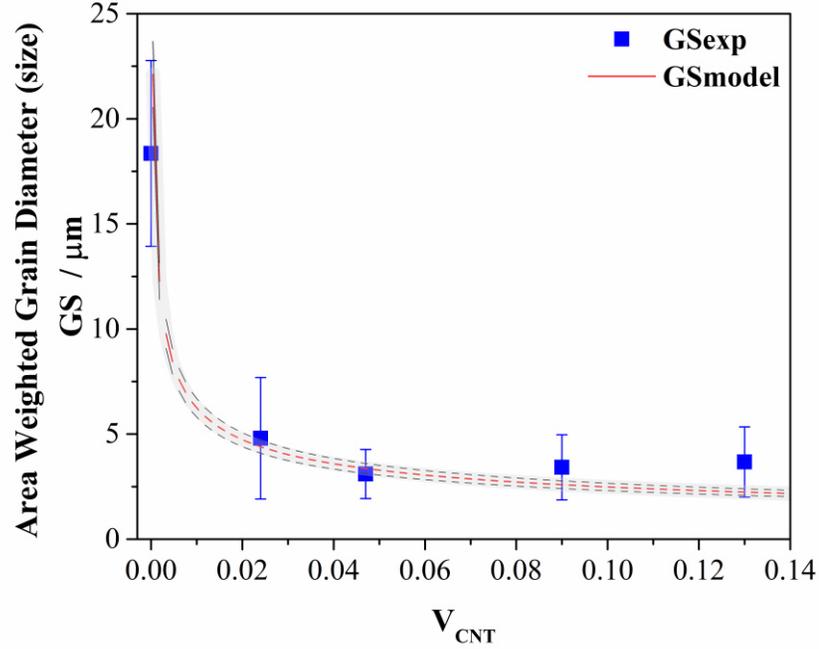

**Figure 6 Mean grain size of studied composites after HUP as a function of the reinforcement volume fraction ($V_{CNT}$). Dashed curves represent the empirical Zener pinning model with respective upper and lower bounds (grey envelope).**

Accordingly, and rearranging equation 8:

$$\varepsilon_{eq} \geq \frac{1.143 V_V^{\left(-\frac{1}{15}\right)}}{D} \left[-\frac{1}{ln(1-V_V)}\right]^{\left(\frac{1}{3}\right)} \quad (9)$$

According to equation 9, there is a minimum strain $\varepsilon^{hom}$ that should be applied in order to obtain a homogenous distribution of CNT. It can be thus inferred that for increasing CNT volume fraction $V_{CNT}$ a lower strain should be applied in order to obtain a homogeneous distribution of CNT after HPT. This corresponds well with the literature, where, for lower $V_{CNT}$, inhomogeneity rises due to the presence of reinforcement-depleted regions [23,39]. Table 1 shows the results for $\varepsilon^{hom}$, according to equation 9, for the different CNT compositions used in this work.

**Table 1 Values of equivalent strain required for achieving a homogenous CNT distribution in MMC processed by HPT.**

| $V_{CNT}$ | $\varepsilon^{hom}$ |
|---|---|
| 0.024 | 126.5 |
| 0.047 | 96.31 |
| 0.09 | 73.71 |
| 0.13 | 63.16 |

In the case of the samples with 1wt. %CNT ($V_{CNT}$ = 0.047) $\varepsilon^{hom} = 96.31$, which is within the range 84 < ε < 100 previously found experimentally, for which these samples were found to be homogenously distributed according to the NND distribution function. These homogenously distributed samples presented values of region homogeneity $H_R \geq 60$ %. It can thus be inferred, that a $H_R$ over 60% is an empirical lower bound, which is necessary but not sufficient to unequivocally identify a homogeneous random distribution of the second phase. It is therefore unavoidable the utilization of a complementary evaluation, as proposed here, which additionally evaluates the distribution of the mean spatial inter-particle distance. Furthermore, this approach provides information valuable for the optimization of the processing parameters by HPT of MMC.

**Conclusions**

The distribution homogeneity of CNT agglomerates in MMC processed by high-pressure torsion is evaluated by the combination of the region homogeneity parameter $H_{RO}$ and the nearest neighbor distribution function $G(r)$. This methodology can be used as a thorough evaluation of the distribution homogeneity of second phases in MMC. Furthermore, CNT agglomerates debond by relative sliding between CNT during HPT, which differs significantly to the debonding-mechanism in ceramic particles. Finally, a model that predicts the minimum equivalent strain required for a homogenous second phase distribution during HPT is developed, which correlates well with the experimental data and provides a basis for the optimization of the processing parameters of MMC by severe plastic deformation.

**Acknowledgements**

K. Aristizabal wishes to thank the German Academic Exchange Service (DAAD) for their financial support. S. Suarez and K. Aristizabal gratefully acknowledge the financial support from DFG (Grant: SU911/1-1). A. Katzensteiner and A. Bachmaier gratefully acknowledge the financial support by the Austrian Science Fund (FWF): I2294-N36.